\documentclass{easychair}

\usepackage{latexsym}

\usepackage{amsthm}
\usepackage{amsfonts}

\newcommand{\xf}[1]{Figure~\ref{#1}}

\newcommand{\xs}[1]{Section~\ref{#1}}

\newcommand{\gee}{{GEE\index{GEE}}}

\newcommand{\gipsy}{{GIPSY\index{GIPSY}}}

\newcommand{\gipl}{{GIPL\index{GIPL}}}

\newcommand{\lucid}{{Lucid\index{Lucid}}}

\newcommand{\lucx}{{Lucx\index{Lucx}}}

\newcommand{\java}{{Java\index{Java}}}

\newcommand{\api}[1]{\texttt{#1}\index{API!#1}}

\newcommand{\lucidL}[1]{{$\mathit{Lucid}$}($L$) }

		{}

\def\myvert{\raise 2.27pt \hbox{\vrule depth 0pt height 8pt width 0.2mm}}
\def\myarrow{\hspace*{0.43mm}%
             \raise 2.29pt\hbox{\vrule depth 0pt height 8pt width 0.16mm}%
             \hspace*{-0.32mm}%
             $\longrightarrow$
             \ %
             }

\lstdefinestyle{codeStyle}
{
	language=Java,
	frame=single,  
	basicstyle=\footnotesize,
	captionpos=b,
	showstringspaces=false,
	showspaces=false,
	extendedchars=true,
	linewidth=1\linewidth,
	breaklines=true,
	float=htb!  
}

\theoremstyle{definition} \newtheorem{xtdef}{Definition}
\theoremstyle{definition} \newtheorem{syntaxrule}{Syntax Rule} 

\theoremstyle{definition} \newtheorem{xteg}{Example}

\begin{document}

%
%

\title{Complete Context Calculus Design and Implementation in GIPSY}
\titlerunning{Complete Context Calculus Design and Implementation in GIPSY}






%
%
%

\author{Xin Tong, Joey Paquet, Serguei A. Mokhov\\
1455 De Maisonneuve Blvd. W.\\
Computer Science and Software Engineering\\
Concordia University\\
Montreal, Quebec, Canada\\
H3G 1M8\\
Email \url{{x_ton,paquet,mokhov}@cse.concordia.ca}}

\authorrunning{Tong, Paquet, and Mokhov}

%
%



\maketitle

%
%

\begin{abstract}
This paper presents the integration into the {\gipsy} of {\lucx}'s context calculus
defined in Wan's PhD thesis. We start by defining different types of tag sets,
then we explain the concept of context, the types of context and the context
calculus operators. Finally, we present how context entities have been abstracted
into Java classes and embedded into the {\gipsy} system.
\end{abstract}

%
%

\section{Introduction}

Lucid~\cite{lucid85,lucid95,nonprocedural-iterative-lucid-77,lucid76,lucid77} represents a family of intensional programming languages that has several dialects all sharing a generic counterpart, which we call the Generic Intensional Programming Language ({\gipl})~\cite{paquetThesis,aihuawu02,chunleiren02,mokhovmcthesis05}. The {\gipl} is a functional programming language whose semantics was defined according to Kripke's possible worlds semantics~\cite{kripke69}. Following this semantics, the notion of context is a core concept, as the evaluation of expressions in intensional programming languages relies on the implicit context of utterance~\cite{paquetThesis}. In earlier versions of {\lucid}, contexts could not be explicitly defined or used in expressions, nor used as first-class values in the language. A new dialect of {\lucid}, which is called {\lucx} (Lucid Enriched With Context) was introduced by Wan~\cite{wanphd06}. {\lucx} embraced the idea of context as first-class value and it also had a collection of context calculus operators defined, coalesced into a well-defined context calculus. However, the operational details of integrating {\lucx} into the GIPSY have not yet been defined, so these latest very important results are not integrated in our operational system. 

\paragraph*{Problem Statement}

In her PhD thesis, Wan has set the basis of a context calculus and demonstrated how it could be integrated into the existing implementation of the GIPSY through the expression of context calculus operators as {\lucid} functions, and the simulation of contexts using {\lucid} finite streams. Such an implementation, though it provided a nice proof of concept, would eventually lead to a notoriously inefficient implementation. What we need is to fully integrate the context calculus into the syntax of the GIPL, as well as to integrate its semantics into the run-time system. Achieving this would bring forth the first intensional language implemented to include contexts as first class value in its syntax and semantics.  

\paragraph*{Proposed Solution}

Based on Wan's theory and the current architecture of the {\gipsy} framework~\cite{mokhovmcthesis05,gipsy2005} we refined and implemented the context calculus including the new syntactical constructs required for the language to be more expressive in terms of implicit and explicit context manipulation, and we embedded a context data type together with the corresponding tag set types into the GIPSY type system~\cite{gipsy-type-system-lpar08}. The introduction of such new constructs required adaptive modifications to our existing implementation, which are described in this paper.

\section{Tag Sets}
\label{tagsets}

A context is essentially a relation between dimensions and tags, the latter being indexes used to refer to points in the context space defined over these dimensions. In {\lucx}, such a relation is represented using a collection of \verb|<dimension:tag>| pairs~\cite{wanphd06}. In such a pair, the current position of the dimension is marked by the tag value, while properties of the tags, such as what are valid tags in this dimension, are bound to the dimension they index. When a context is declared, a semantic check should be performed to determine whether a tag is valid in the dimension it is used. Therefore, we introduce the notion of a {\em tag set}, as a collection of all possible tags attached to a particular dimension, i.e. we introduce the notion of {\it tag types}. 

In earlier versions of Lucid programming languages, the tag set was assumed to be the ordered infinite set of natural numbers, and was never explicitly declared as such. However, as we explore more domains of application, natural numbers can no longer represent tag values sufficiently. For example, assume that we want to compute the gravity of certain planets in the solar system. We could define \verb|planet| as our dimension; \verb|#planet| returns the current tag in the \verb|planet| dimension. We use the square brackets notation \texttt{[planet:\#planet]} to represent a {\it simple context}~\cite{wanphd06}, as a collection of \verb|<dimension:tag>| pairs. The result of the program should be a stream of \verb|gravity| values. The evaluation of specific values in this stream depends on the specific context of utterance, such as \verb|gravity@[planet:3]|.

If we set our focus onto the planets inside the current solar system, then up to our knowledge now, there is a finite number of planets, i.e. the tag set for the \verb|planet| dimension is \verb|[1..8]|. We could also make the tag set of this dimension into the set {\tt \{Mercury, Venus, Earth, Mars, Jupiter, Saturn, Uranus, Neptune\}}, where the tags are no longer integers representing the order of proximity to the Sun, but strings representing the names of the planets. Note that such a tag set could still be ordered by the order of proximity to the Sun, as represented here, or alphabetically. If we extend the dimension to all possible planets in the universe, then the number of tags would be infinite, and thus could not be enumerated. Note also that the order defined on the tag set is of importance, as basic operators such as \verb|fby| and \verb|next| rely on an ordered tag set. It should thus be possible to define an order on tag sets, and declaring a tag set as {\it unordered} would then restrict the set of operators applicable to streams defined on a dimensions with an unordered tag set. It is thus clear that the properties of natural numbers set--ordered and infinite--are not sufficient to include all the possibilities for all possible tag set types. Additionally, the tag value can actually be of \api{string} or other types, not only \api{int}. Thus, it is necessary to introduce the keywords ``ordered/unordered'', ``finite/infinite'' to determine the types of tag set associated with dimensions upon declaration. Note that more keywords might also be included in the future, here we only present those to the scope of our knowledge and the current application. Following are the definitions for those keywords when they are used to determine the type of a tag set. As tag sets are in fact sets, we define the following terms as of set theory~\cite{settheory1, settheory2}:

\begin{xtdef}
{\bf Ordered Set:} A set on which a relation $R$ satisfies the following three properties :
\begin{enumerate}
\item Reflexive: For any $a \in S$, we have $aRa$ 
\item Antisymmetric: If $aRb$ and $bRc$, then $a=b$
\item Transitive: If $aRb$ and $bRc$, then $aRc$
\end{enumerate}
\end{xtdef}

\begin{xtdef}
{\bf Unordered Set:} A set which is not ordered is called an unordered set.
\end{xtdef}

\begin{xtdef}
{\bf Finite Set:} A set $I$ is called finite and more strictly, inductive, if there exists a positive integer $n$ such that $I$ contains just $n$ members. The null-set $\emptyset$ is also called finite.
\end{xtdef}

\begin{xtdef}
{\bf Infinite Set:} A set, which is not finite is called an infinite set.
\end{xtdef}

Out of backwards compatibility with previous versions of {\lucid}, we assume that the default tag set is the integers, and its order is as with the order of natural numbers. If other tag sets are to be applied, the programmer must specify them by explicitly specifying and/or enumerating the tag set and its order, as discussed further in this section.

\subsection{Tag Set Types}

In the following sections, the actual types of tag sets are rendered by providing their syntax in {\lucx}'s implementation, followed by the applicability for these syntax rules, then some examples, and finally the implementation of set inclusion routines applicable to all these tag types.

\subsubsection{Ordered Finite Tag Set}

For this type, tags inside the tag set are ordered and finite. Here we use $\mathbb{Z}$ to denote the set of all integers; $S$ to denote the tag set. 
We define $l, u, p, e \in \mathbb{Z}$ as integers to denote the lower boundary ($l$), upper boundary ($u$), step ($p$) and any element ($e$) of the tag set when describing it syntactically.
Also note that $prev(e)$ returns the element previous to the current element under discussion.  
\begin{syntaxrule}
{\tt dimension} $<\!\!id\!\!>$: {\tt ordered} {\tt finite} {\tt\{}$<\!\!string\!\!>,\ldots,<\!\!string\!\!>${\tt\}}
\begin{itemize}
\item All the tag values inside the tag set are enumerated and their order is implicitly defined as the order in which they are enumerated.
\end{itemize}
\end{syntaxrule}

\begin{syntaxrule}
{\tt dimension} $<\!\!id\!\!>$: {\tt ordered} {\tt finite} $\{l\; {\tt to}\; u\}$
    \begin{itemize}
        \item $S \subset \mathbb{Z} = \{e | e-prev(e)=1 \wedge l \leq e \leq u \}$
    \end{itemize}
\end{syntaxrule}

\begin{syntaxrule}
{\tt dimension} $<\!\!id\!\!>$: {\tt ordered} {\tt finite} {\tt\{}$l\; {\tt to}\; u\; {\tt step}\; p${\tt\}}
\begin{itemize}

\item $S \subset \mathbb{Z} = \{e | e-prev(e)=p \wedge l \leq e \leq u \wedge p>0 \}$

\item $S \subset \mathbb{Z} = \{e | e-prev(e)=p \wedge u \leq e \leq l \wedge p<0 \}$

\end{itemize}
\end{syntaxrule}

\begin{xteg}
The following examples correspond to the syntactic expressions listed above, respectively. 
\begin{itemize}
\item \texttt{dimension d : ordered finite \{rat, bull, tiger, rabbit\}}
\item \texttt{dimension d : ordered finite \{1 to 100\}}
\item \texttt{dimension d : ordered finite \{2 to 100 step 2\}}
\end{itemize}
\end{xteg}

\paragraph{Set Inclusion}

\begin{itemize}
\item
If it is in the first format of expression, then set inclusion returns true if and only if the given parameter is equal to one of the tag values inside the tag set as extensionally enumerated.
\item
If the tag set is declared using the second format, then set inclusion returns true if and only if the given parameter is greater than or equal to the lower boundary and smaller than or equal to the upper boundary.
\item
If the third expression is applied, then set inclusion returns true if the given parameter $para$ is greater than or equal to the lower boundary, and smaller than or equal to the upper boundary, if the step is possitive; or smaller than or equal to the lower boundary and greater than or equal to the upper boundary if the step is negative; and that $((para-l)\;{\mathtt{mod}}\;p) = 0$ in both cases.
\end{itemize}

\subsubsection{Ordered Infinite Tag Set}

For this type, tags inside the tag set are ordered and infinite. Since the tag set is infinite, it cannot be enumerated. For now, we only consider subsets of integers. Note, in what follows \texttt{INF-} and \texttt{INF+} stand for minus infinity ($-\infty$) and plus infinity ($+\infty$) respectively.

\begin{syntaxrule}
{\tt dimension} $<\!\!id\!\!>$: {\tt ordered} {\tt infinite} {\tt\{}$l$ {\tt to INF+}{\tt\}}
\begin{itemize}
\item $S \subset \mathbb{Z} = \{e | e-prev(e)=1 \wedge l \leq e \}$
\end{itemize}
\end{syntaxrule}

\begin{syntaxrule}
{\tt dimension} $<\!\!id\!\!>$: {\tt ordered} {\tt infinite} {\tt\{}$l$ {\tt to} {\tt INF+} {\tt step} $p${\tt\}}

\begin{itemize}
\item $S \subset \mathbb{Z} = \{e | e-prev(e)=p \wedge l \leq e \wedge p>0 \}$

\end{itemize}
\end{syntaxrule}

\begin{syntaxrule}
{\tt dimension} $<\!\!id\!\!>$: {\tt ordered} {\tt infinite} {\tt\{}{\tt INF-} {\tt to} $u${\tt\}}
\begin{itemize}
\item $S \subset \mathbb{Z} = \{e | e-prev(e)=1 \wedge e \leq u \}$
\end{itemize}
\end{syntaxrule}

\begin{syntaxrule}
{\tt dimension} $<\!\!id\!\!>$: {\tt ordered} {\tt infinite} {\tt\{}{\tt INF-} {\tt to} $u$ {\tt step} $p${\tt\}}
\begin{itemize}
\item $S \subset \mathbb{Z} = \{e | e-prev(e)=p \wedge e \leq u \wedge p>0 \}$
\end{itemize}
\end{syntaxrule}

\begin{syntaxrule}
{\tt dimension} $<\!\!id\!\!>$: {\tt ordered} {\tt infinite} {\tt\{}{\tt INF- to INF+\}}
\begin{itemize}
\item This represents the whole stream of integers, from minus infinity to plus infinity.
\end{itemize}
\end{syntaxrule}

\noindent
Note that the default tag set is $\mathbb{N}^+$, which is also within this type. Either by leaving the tag set declaration part empty or specifying {\tt\{0 to INF+\}}, they both refer to the set of natural numbers.

\begin{xteg}
The following examples correspond to the syntactic expressions listed above, respectively.
\begin{itemize}
\item \texttt{dimension d : ordered infinite \{2 to INF+\}}
\item \texttt{dimension d : ordered infinite \{2 to INF+ step 2\}}
\item \texttt{dimension d : ordered infinite \{INF- to 100\}}
\item \texttt{dimension d : ordered infinite \{INF- to 100 step 2\}}
\item \texttt{dimension d : ordered infinite \{INF- to INF+\}}
\end{itemize}
\end{xteg}

\paragraph{Set Inclusion}

Although we call this type of set `infinite', in the actual implementation, there should be
a way to handle this `infinity' to make it `infinite' allowed by the available storage resources.
For now we only consider \api{Integer} as the type for a tag value, thus the infinity is
actually represented by either \api{Integer.MIN\_VALUE} of {\java} for minus
infinity or \api{Integer.MAX\_VALUE} for plus infinity. The set inclusion method is defined
and implemented as the following:

\begin{itemize}
\item
If the first expression is applied: then set inclusion method returns true if and only if
the given parameter is greater than or equal to the lower boundary and less than
or equal to \api{Integer.MAX\_VALUE}.
\item
If it is in the second format: then set inclusion method returns true if and only if
the given parameter $para$ is greater than or equal to the lower boundary and less than or
equal to \api{Integer.MAX\_VALUE} and that $((para-l)\;{\mathtt{mod}}\;p) = 0$.
\item
If the third expression is used: then set inclusion method returns true if and only if
the given parameter is less than or equal to the upper boundary and greater than or equal
to \api{Integer.MIN\_VALUE}.
\item
If it is declared in the forth format: then set inclusion method returns true if and only
if the given parameter $para$ is less than or equal to the upper boundary and greater than or
equal to \api{Integer.MIN\_VALUE} and that $((u-para)\;{\mathtt{mod}}\;p) = 0$.
\item
Finally, if it is in the fifth expression: then the set inclusion method returns true if and
only if the given parameter is greater than or equal to \api{Integer.MIN\_VALUE} and
less than or equal to \api{Integer.MAX\_VALUE}.
\end{itemize}

\subsubsection{Unordered Finite Tag Set}

Tags of this type are unordered and finite.
\begin{syntaxrule}
{\tt dimension} $<\!\!id\!\!>$: {\tt unordered} {\tt finite} {\tt\{}$<\!\!string\!\!>,\ldots,<\!\!string\!\!>${\tt\}}
\end{syntaxrule}

\begin{xteg}
The following example correspond to the syntactical expression above. 
\begin{itemize}
\item \texttt{dimension d: unordered finite \{red, yellow, blue\}}
\end{itemize}
\end{xteg}

\paragraph{Set Inclusion}

The set inclusion method returns true if and only if the given parameter is equal to one of the tag values inside the tag set.

\subsubsection{Unordered Infinite Tag Set}

Tags of this type are unordered and infinite.

\begin{syntaxrule}
{\tt dimension} $<\!\!id\!\!>$: {\tt unordered} {\tt infinite} {\tt\{}$<\!\!E\!\!>${\tt\}}
\end{syntaxrule}

The $<\!\!E\!\!>$ could be either intensional functions generating unordered infinite elements or imperative procedures such as Java methods to generate such elements. See the example below for a discussion. 

\begin{xteg}
Assume that we have a device to collect sound waves and it has a software
interface to computers. And we have a \api{getWave()} method defined somewhere,
which returns all the sound waves that can be detected by the device. If we want
to set the device working `infinitely' (ideally) in the sea in order to filter
the sound waves of sperm whales to keep track of their conditions, we would
define our tag set as:

\vspace{3mm}
\texttt{ \{ while(true) \{ getWave(); \} \}}
\vspace{3mm} 

As this type of tag set is unordered and infinite, it's impossible to enumerate all the tag values
in the tag set. The programmer has to provide a function to define all the possible tag values.
Since some random number generator functions can also be considered valid for this type, the set
inclusion can only be determined by the type of tag value. For example, if the random function
generates only integers, then a tag value specified as any other type in the program should
not be inside the tag set.
\end{xteg}

\section{Context Calculus}

Context calculus operators are a set of operators performed on contexts. All the following definitions are recited from Wan's PhD thesis.
We present here only an overview of the theory underlying the notion of context and its calculus for the unaware readers. For a complete description please refer to~\cite{wanphd06}.

\begin{xtdef}
{\bf Context:} A context $c$ is a finite subset of the relation: $c \subset \{(d, x) | d \in DIM \wedge x \in T\}$, where $DIM$ is the set of all possible dimensions, and $T$ is the set of all possible tags. 
\end{xtdef}

\subsection{Types of Context}
According to \cite{wanphd06}, context can be classified into two categories, which are {\em simple context} and a {\em context set}.

\subsubsection{Simple Context}

A {\it simple context} is a collection of $<\!\!dimension:tag\!\!>$ pairs, where there are no two such pairs having the same dimension component. Conceptually, a simple context represents a point in the context space. A simple context having only one pair of $<\!\!dimension:tag\!\!>$ is called a {\em micro context}. It is the building block for all the context types~\cite{tongxinmcthesis08,gipsy-simple-context-calculus-08}.

\begin{syntaxrule}
${\mathtt{[}}<\!\!E\!\!>:<\!\!E\!\!>,\ldots,<\!\!E\!\!>:<\!\!E\!\!>]$
\end{syntaxrule}

\begin{xteg}
\begin{itemize}
\item \verb|[d:1,e:2]|
\end{itemize}
\end{xteg}

\subsubsection{Context Set}

A context set is a set of simple contexts. Context sets are also
often named {\em non-simple contexts}. Context sets represent regions
of the context space, which can be seen as a set of points in the context space,
considering that the context space is discrete. Formally speaking, a non-simple
context is a set of $<\!\!d:x\!\!>$ mappings that are not defined by a function \cite{gipsywiki}.
The semantics of context set has not been integrated into the {\lucid} programming language,
yet, informally, as a context set can be viewed as a set of simple context,
the semantic rules will apply on each element individually.

\begin{syntaxrule}
$\{[<\!\!E\!\!>:<\!\!E\!\!>,\ldots,<\!\!E\!\!>:<\!\!E\!\!>],\ldots,[<\!\!E\!\!>:<\!\!E\!\!>,\ldots,<\!\!E\!\!>:<\!\!E\!\!>]\}$
\end{syntaxrule}

\begin{xteg}
\begin{itemize}
\item \verb|{[x:3,y:4,z:5],[x:3,y:1,z:5]}|
\end{itemize}

\end{xteg}

\subsection{Context Calculus Operators}

In the following section, we provide the formal definition for the context calculus operators on simple context and context set;
and the algorithm for implementing those operators. The operators are \api{isSubContext}, \api{difference}, \api{intersection}, \api{projection}, \api{hiding}, \api{override}, and \api{union}.

\begin{xtdef}
\api{isSubContext}

\begin{itemize}
\item
If $C_1$ and $C_2$ are simple contexts and every micro context of $C_1$ is also a micro context of $C_2$, then $C_1$ \api{isSubContext} $C_2$ returns true: $C_1=\{m_1,\ldots,m_i\}$ where $m_i$ is any micro context inside $C_1$. If $m_i \in C_2$, then $C_1$ \api{isSubContext} $C_2$ returns true. Note that an empty simple context is the sub-context of any simple context. Also note that as the concept of subset in set theory, $C_1$ could be the proper subset of $C_2$, or $C_1$ could be equal to $C_2$.

\item
If $S_1$ and $S_2$ are context sets and every simple context of $S_1$ is also a simple context of $S_2$, then $S_1$ \api{isSubContext} $S_2$ returns true. $S_1=\{C_1,\ldots,C_i\}$ where $C_i$ is any simple context inside $S_1$. If $C_i \in S_2$, then $S_1$ \api{isSubContext} $S_2$ returns true. Note that an empty context set is the sub-context of any context set. Also note that as the concept of subset in set theory, $S_1$ could be the proper subset of $S_2$, or $S_1$ could be equal to $S_2$.
\end{itemize}

\end{xtdef}

\begin{xteg}
Example for \api{isSubContext} on both simple context and context set.

\begin{itemize}
\item \verb|[d:1,e:2]| \api{isSubContext} \verb|[d:1,e:2,f:3]| = true 
\item \verb|[d:1,e:2]| \api{isSubContext} \verb|[d:1,e:2]| = true
\item $\emptyset$ \api{isSubContext} \verb|[d:1,e:2]|= true
\item \verb|{[d:1,e:2],[f:3]}| \api{isSubContext} \verb|{[d:1,e:2],[f:3],[g:4]}| = true
\item \verb|{[d:1,e:2],[f:3]}| \api{isSubContext} \verb|{[d:1,e:2],[f:3]}| = true.
\end{itemize}
\end{xteg}

\begin{lstlisting}[
    label={list:isSubcontextSimpleContext},
    caption={Algorithm for implementing \api{isSubContext} on simple context},
    style=codeStyle
    ]
boolean isSubContext(SimpleContext c1, SimpleContext c2)
{
  if(c1.size == 0)
    return true;
  else{
     boolean flag;
     for(int i = 0; i < c1.size; i++){
       flag=false;
       for(int j = 0; j < c2.size; j++){
         if(c2.micro_context(j) == c1.micro_context(i)){
           flag=true;
           break;
         }
       }
        if(flag==false)
          break;
      }      
      return flag;
  }
}
\end{lstlisting}

\begin{lstlisting}[
    label={list:isSubcontextContextSet},
    caption={Algorithm for implementing \api{isSubContext} on context set},
    style=codeStyle
    ]
boolean isSubContext(ContextSet s1, ContextSet s2)
{
  if(s1.size == 0)
    return true;
  else{
    boolean flag;
    for(int i = 0; i < s1.size; i++){
      flag=false;
      for(int j = 0; j < s2.size; j++){
        if(s2.simple_context(j)==s1.simple_context(i)){
          flag=true;
          break;
        }
      }
        if(flag == false)
          break;
     }     
      return flag;
  }
}
\end{lstlisting}
                   

\begin{xtdef}
\api{difference:}

\begin{itemize}
\item
If $C_1$ and $C_2$ are simple contexts, then $C_1$ \api{difference} $C_2$
returns a simple context that is the collection of all micro contexts which
are members of $C_1$, but not members of $C_2$: $C_1=\{m_1,\ldots\,m_i\}$
where $m_i$ is any micro context inside $C_1$.
$C_1$ \api{difference} $C_2=\{m_i| m_i \notin C_2\}$.
Note that if $C_1$ \api{isSubContext} $C_2$ is true, then the returned
simple context should be the empty context. Also note that it is
valid to ``differentiate'' two simple contexts that have no common micro
context; the returned simple context is simply $C_1$.  

\item
If $S_1$ and $S_2$ are context sets, this operator returns a context set $S$,
where every simple context $C \in S$ is computed as $C_1$ \api{difference} $C_2$,
$C_1 \in S_1$, $C_2 \in S_2$: $S = S_1$ \api{difference} $S_2 = \{C_1$ \api{difference} $C_2 | C_1 \in S_1 \wedge C_2 \in S_2$\}.
\end{itemize}

\end{xtdef}

\begin{xteg}
Example for \api{difference} on both simple context and context set.
\begin{itemize}

\item \verb|[d:1,e:2]| \api{difference} \verb|[d:1,f:3]| = \verb|[e:2]|
\item \verb|[d:1,e:2]| \api{difference} \verb|[d:1,e:2,f:3]| = $\emptyset$
\item \verb|[d:1,e:2]| \api{difference} \verb|[g:4,h:5]| = \verb|[d:1,e:2]|
\item \verb|{[d:1,e:2,f:3],[g:4,h:5]}| \api{difference} \verb|{[g:4,h:5],[e:2]}| = \\
\verb|{[d:1,e:2,f:3],[d:1,f:3],[g:4,h:5]]|
\end{itemize}

\end{xteg}

\begin{lstlisting}[
    label={list:differenceSimpleContext},
    caption={Algorithm for implementing \api{difference} on simple context},
    style=codeStyle
    ]
SimpleContext difference(SimpleContext c1, SimpleContext c2){
  SimpleContext result=c1.clone(); 
    for(int i = 0; i < c1.size; i++){
      for(int j = 0; j < c2.size; j++){      
        if(c2.micro_context(j)==c1.micro_context(i)){
          result.remove(c1.micro_context(i));
        }
      }
    }
  return result;
}
\end{lstlisting}

\begin{lstlisting}[
    label={list:differenceContextSet},
    caption={Algorithm for implementing \api{difference} on context set},
    style=codeStyle
    ]
ContextSet difference(ContextSet s1, ContextSet s2){
  ContextSet result;
  for(int i = 0; i < s1.size; i++){
    for(int j = 0; j < s2.size; j++){
      SimpleContext tempResult=difference(s1.simple_context(i), s2.simple_context(j));
      if(tempResult.size != 0)
        result.add(tempResult);
    }
  }
  return result;
}
\end{lstlisting}


\begin{xtdef}
\api{intersection}

\begin{itemize}
\item
If $C_1$ and $C_2$ are simple contexts, then $C_1$ \api{intersection} $C_2$
returns a new simple context, which is the collection of those micro contexts
that belong to both $C_1$ and $C_2$:
$C_1=\{m_1,\ldots\,m_i\}$ where $m_i$ is any micro context inside $C_1$:
$C_1$ \api{intersection} $C_2=\{m_i| m_i\in C_1\wedge m_i\in C_2\}$.
Note that if $C_1$ and $C_2$ have no common micro contexts, the result
is an empty simple context.

\item
If $S_1$ and $S_2$ are context sets, then the resulting intersection set $S = S_1$ \api{intersection} $S_2 = \{C_1$ \api{intersection} $C_2 | C_1 \in S_1 \wedge C_2 \in S_2 \}$
\end{itemize} 
\end{xtdef}

\begin{xteg}
Example for \api{intersection} on both, simple context and context set:
\begin{itemize}
\item \verb|[d:1,e:2]| \api{intersection} \verb|[d:1]| = \verb|[d:1]|
\item \verb|[d:1,e:2]| \api{intersection} \verb|[g:4,h:5]| = $\emptyset$
\item \verb|{[d:1,e:2,f:3],[g:4,h:5]}| \api{intersection} \verb|{[g:4,h:5],[e:2]}| = \\
\verb|{[e:2],[g:4,h:5]}|
\end{itemize}

\end{xteg}

\begin{lstlisting}[
    label={list:intersectionSimpleContext},
    caption={Algorithm for implementing \api{intersection} on simple context},
    style=codeStyle
    ]
SimpleContext intersection(SimpleContext c1, SimpleContext c2){
  return difference(c1, difference(c1, c2));
}
\end{lstlisting}

\begin{lstlisting}[
    label={list:intersectionContextSet},
    caption={Algorithm for implementing \api{intersection} on context set},
    style=codeStyle
    ]
ContextSet intersection(ContextSet s1, ContextSet s2){
  ContextSet result;
  for(int i = 0; i < s1.size; i++){
    for(int j = 0; j < s2.size; j++)
    {
      SimpleContext tempResult=intersection(s1.simple_context(i), s2.simple_context(j));
      if(tempResult.size != 0)
        result.add(tempResult);
    }
  }
  return result;
}
\end{lstlisting}

\begin{xtdef}
\api{projection}: 

\begin{itemize}
\item
If $C$ is a simple context and $D$ is a set of dimensions, this operator
filters only those micro contexts in $C$ that have their dimensions in
set $D$. $C$ \api{projection} $D = \{m | m \in C \wedge dim(m) \in D\}$.
Note that if there's no micro context having the same dimension as in the
dimension set, the result would be an empty simple context.
$dim(m)$ returns the dimension of micro context $m$.

\item
The projection of a context set and a dimension set is a context set,
which is a collection of all the simple contexts project the dimension set.
If $S$ is a context set, $D$ is a dimension set; $S$ \api{projection} $D=\{n | n$ = $C$ \api{projection} $D \wedge C \in S$\}.
Note that if there's no common dimension in every simple context and the
dimension set, the result is an empty context set.
\end{itemize}
\end{xtdef}

\begin{xteg}
Example of \api{projection} on both simple context and context set:

\begin{itemize}
\item \verb|[d:1,e:2,f:3]| \api{projection} \verb|{d,f}| = \verb|[d:1,f:3]|
\item \verb|{[d:1,e:2,f:3],[g:4,h:5],[f:4]}| \api{projection} \verb|{e,f,h}| =\\
\verb|{[e:2,f:3],[h:5],[f:4]}|
\end{itemize}
\end{xteg}

\begin{lstlisting}[
    label={list:projectionSimpleContext},
    caption={Algorithm for implementing \api{projection} on simple context},
    style=codeStyle
    ]
SimpleContext projection(SimpleContext c, DimensionSet dimSet){
  SimpleContext result;
  for(int i = 0; i < dimSet.size; i++){
    for(int j = 0; j < c.size; j++){
      if(c.micro_context(j).dimension == dimSet.dimension(i))
        result.add(c.micro_context(j));
    }
  }
  return result;
}
\end{lstlisting}

\begin{lstlisting}[
    label={list:projectionContextSet},
    caption={Algorithm for implementing \api{projection} on context set},
    style=codeStyle
    ]
ContextSet projection(ContextSet s, DimensionSet dimSet){
  ContextSet result;
  for(int i = 0; i < s.size; i++){
    SimpleContext tempResult=projection(s.simple_context(i), dimSet);
    if(tempResult.size != 0)
      result.add(tempResult);
  }
  return result;
}
\end{lstlisting}

\begin{xtdef}
\api{hiding}:

\begin{itemize}
\item
If $C$ is a simple context and $D$ is a dimension set, this operator is to remove
all the micro contexts in $C$ whose dimensions are in $D$: $C$ \api{hiding} $D = \{m | m \in C \wedge dim (m) \notin D\}$.
Note that $C$ \api{projection} $D \bigcup C$ \api{hiding} $D = C$.

\item
For context set $S$, and dimension set $D$, the \api{hiding} operator constructs
a context set $S'$ where $S'$ is obtained by \api{hiding} each simple context
in $S$ on the dimension set $D$: $S' = S$ \api{hiding} $D = \{C$ \api{hiding} $D | C \in S$\}.
\end{itemize}
\end{xtdef}

\begin{xteg}
Example for \api{hiding} on both simple context and context set:

\begin{itemize}
\item \verb|[d:1,e:2,f:3]| \api{hiding} \verb|{d,e}| = \verb|[f:3]|
\item \verb|[d:1,e:2,f:3]| \api{hiding} \verb|{g,h}| = \verb|[d:1,e:2,f:3]|
\item \verb|[d:1,e:2,f:3]| \api{hiding} \verb|{d,e,f}| = $\emptyset$
\item \verb|{[d:1,e:2,f:3],[g:4,h:5],[e:3]}| \api{hiding} \verb|{d,e}| = \verb|{[f:3],[g:4,h:5]}|
\end{itemize}
\end{xteg}

\begin{lstlisting}[
    label={list:hidingSimpleContext},
    caption={Algorithm for implementing \api{hiding} on simple context},
    style=codeStyle
    ]
SimpleContext hiding(SimpleContext c, DimensionSet dimSet){
  return(difference(c, projection(c, dimSet)));
}
\end{lstlisting}

\begin{lstlisting}[
    label={list:hidingContextSet},
    caption={Algorithm for implementing \api{hiding} on context set},
    style=codeStyle
    ]
ContextSet hiding(ContextSet s, DimensionSet dimSet){
  ContextSet result;
  for(int i = 0; i < s.size; i++){
    SimpleContext tempResult=hiding(s.simple_context(i), dimSet);
    if(tempResult.size != 0)
      result.add(tempResult);
  }
  return result;
}
\end{lstlisting}

\begin{xtdef}
\api{override}:

\begin{itemize}
\item
If $C_1$ and $C_2$ are simple contexts, then $C_1$ \api{override} $C_2$ returns a new simple
context $C$, which is the result of the conflict-free union of $C_1$ and $C_2$, as defined below:
$C = C_1$ \api{override} $C_2 = \{m | (m \in C_1 \wedge dim(m) \notin dim(C_2)) \vee m \in C_2\}$.

\item
For every pair of context sets $S_1$, $S_2$, this operator returns a set of contexts $S$,
where every context $C \in S$ is computed as $C_1$ \api{override} $C_2$;
$C_1 \in S_1$, $C_2 \in S_2$: $S = S_1$ \api{override} $S_2 = \{C_1$ \api{override} $C_2 | C_1 \in  S_1 \wedge C_2 \in S_2$\}.
\end{itemize}
\end{xtdef}
\begin{xteg}
Example of \api{override} on both simple context and context set:

\begin{itemize}
\item \verb|[d:1,e:2,f:3]| \api{override} \verb|[e:3]| = \verb|[d:1,e:3,f:3]|
\item \verb|[d:1,e:2,f:3]| \api{override} \verb|[e:3,g:4]| = \verb|[d:1,e:3,f:3,g:4]|
\item \verb|{[d:1,e:2],[f:3],[g:4,h:5]}| \api{override} \verb|{[d:3],[h:1]}| =\\
			\verb|{[d:3,e:2],[d:1,e:2,h:1],[f:3,d:3],|\\
			\verb|[f:3,h:1],[g:4,h:5,d:3],[g:4,h:1]}|
\end{itemize}
\end{xteg}

\begin{lstlisting}[
    label={list:overrideSimpleContext},
    caption={Algorithm for implementing \api{override} on simple context},
    style=codeStyle
    ]
SimpleContext override(SimpleContext c1, SimpleContext c2){
  SimpleContext result;
  boolean flag=false; 
  //keep the micro contexts whose dimensions is in c2, but not in c1
  SimpleContext uniqueMCInC2=c2.clone();
  for(int i = 0; i < c1.size; i++){
    for(int j = 0; j < c2.size; j++){
      if(c1.micro_context(i).dimension == c2.micro_context(j).dimension){
        flag=true;
        result.add(c2.micro_context(j));
        uniqueMCInC2.remove(c2.micro_context(j));
        }
    }
    if(flag == false)
      //Add the micro contexts in c1 with unique dimensions
      result.add(c1.micro_context(i));
      flag=false;
  }
  for(int k = 0; k < uniqueMCInC2.size; k++){
    //Add the micro contexts in c2 with unique dimensions
    result.add(uniqueMCInC2.micro_context(k)); 
  }
  return result;
}
\end{lstlisting}

\begin{lstlisting}[
    label={list:overrideContextSet},
    caption={Algorithm for implementing \api{override} on context set},
    style=codeStyle
    ]
ContextSet override(ContextSet s1, ContextSet s2){
  ContextSet result;
  for(int i = 0; i < s1.size; i++){
    for(int j = 0; j < s2.size; j++){
      SimpleContext tempResult=override(s1.simple_context(i), s2.simple_context(j));
      if(tempResult.size != 0)
        result.add(tempResult);
    }
  }
  return result;
}
\end{lstlisting}

\begin{xtdef}
\api{union}:

\begin{itemize}
\item
If $C_1$ and $C_2$ are simple contexts, then $C_1$ \api{union} $C_2$ returns a
new simple context $C$, for every micro context $m$ in $C$: $m$ is an element
of $C_1$ or $m$ is an element of $C_2$: $C_1$ \api{union} $C_2 = \{m| m \in C_1 \vee m \in C_2 \wedge m \notin C_1 \}$.
Note that if there is at least one pair of micro contexts in $C_1$ and $C_2$ sharing
the same dimension and these two micro contexts are not equal then the result is a non-simple context, which can be translated into context set: For a non-simple context $C$, we construct the set $Y = \{y_d = C$ \api{projection} $\{d\} | d \in dim(C)\}$. Denoting the elements of set $Y$ as $y_1, \ldots , y_p$, we construct the set $S(C)$ of simple contexts:
$S(C) = \{m_1$ \api{override} $m_2$ \api{override} \ldots \api{override} $m_p | m_1 \in y1 \wedge m_2 \in y_2 \wedge \ldots m_p \in y_p\}$,
The non-simple context is viewed as the set $S(C)$. It is easy to see that
$S(C) = \{s \in S | dim(s) = dim(C) \wedge s \subset C\}$

\item
As described earlier for the \api{union} operator performing on simple contexts,
the result could be a non-simple context. If we simply compute union for each pair
of simple context inside both context sets, the result may be a set of sets,
in other words, {\em higher-order sets}~\cite{wanphd06}. Due to unnecessary semantic complexities, we should
avoid the occurrence of such sets, thus we define the \api{union} of two context sets
as following to eliminate the possibility of having a higher-order set.
If $C_1$ and $C_2$ are context sets, then $C = C_1$ \api{union} $C_2$ is computed as follows:
$D_1 = \{dim(m) \wedge m\in C_1\}, D_2 = \{dim(m) \wedge m \in C_2\}, D_3 = D_1 \bigcap D_2.$

\begin{enumerate}
\item Compute $X_1: X_1 = \{m_i \bigcup (m_j \; \api{hiding} \; D_3) \wedge m_i \in C1 \wedge m_j \in C_2\}$
\item Compute $X_2: X_2 = \{m_j \bigcup (m_i \; \api{hiding} \; D_3) \wedge m_i \in C1 \wedge m_j \in C_2\}$
\item The result is: $C = X_1 \bigcup X_2$ 
\end{enumerate}

\end{itemize}
\end{xtdef}

\begin{xteg}
Example of \api{union} on both simple context and context set:

\begin{itemize}
\item \verb|[d:1,e:2]| \api{union} \verb|[f:3,g:4]| = \verb|[d:1,e:2,f:3,g:4]|
\item \verb|[d:1,e:2]| \api{union} \verb|[d:3,f:4]| = \verb|[d:1,d:3,f:4]| $\Leftrightarrow$ \verb|{[d:1,f:4],[d:3,f:4]}|
\item \verb|{[d:1,e:2],[g:4,h:5]}| \api{union} \verb|{[g:4,h:5],[e:3]}| =\\
			\verb|{[d:1,e:2],[g:4,h:5],[g:4,h:5,d:1],[e:3,d:1],[e:3]}|
\end{itemize}

\end{xteg}

\begin{lstlisting}[
    label={list:unionSimpleContext},
    caption={Algorithm for implementing \api{union} on simple context},
    style=codeStyle
    ]
Context union(SimpleContext c1, SimpleContext c2){ 
  //Note that the return type is generic 
  //Assume [f:1, e:1, d:2] union [e:2, d:1, t:4]
  SimpleContext result1;
  ContextSet result2;
  boolean isContextSet=false;
  for(int i = 0; i < c1.size; i++){
    for(int j = 0; j < c2.size; j++){
      if(c1.micro_context(i).dimension == c2.micro_context(j).dimension && c1.micro_context(i) != c2.micro_context(j)){ 
        //[e:1, d:2] union [e:1] is a simple context: [e:1, d:2]
        isContextSet=true;
        break;
      }
    }
    if(isContextSet == true)
      break;
  }
  if(isContextSet == false){ 
    //No dimension is common, result is the combination c1 and c2.
    for(int i = 0; i < c1.size; i++){
      result1.add(c1.micro_context(i));
    }
    for(int j = 0; j < c2.size; j++){
      result1.add(c2.micro_context(j));
    }
    //remove duplicates e.g. [e:1, e:1, d:1] becomes [e:1, d:1]
    result1.removeDuplicateContext(); 
    return result1;
  }
  else{ 
    //There are common dimensions, the result is a non-simple context 
    //A function is called to translate it into a context set
    result2=translateContextSet(c1, c2); 
    result2.removeDuplicateContext();
    return result2;
  } 
}
\end{lstlisting}

\begin{lstlisting}[
    label={list:helpingMethodForUnionSimpleContext_translateContextSet},
    caption={Algorithm for implementing helping method \api{translateContextSet} for \api{union} on simple context},
    style=codeStyle
    ]
ContextSet translateContextSet(SimpleContext pC1, SimpleContext pC2){
  //collection of micro contexts in pC1 having common dimensions
  Vector commonMC1, commonMC2;
  //collection of micro contexts in pC1 having no common dimension
  Vector uniqueMC1, uniqueMC2;
  //[e:1, e:2] or [d:1, d:2]
  Vector interMicroContext_i;
  //collection of interMicroContext_i: {[e:1, e:2],[d:1, d:2]} 
  Vector interMicroContext; 
  //{[e:1, d:1], [e:1, d:2], [e:2, d:1], [e:2, d:2]}
  ContextSet commonCombination; 
  //{[f:1, e:1, d:1, t:4]...}
  ContextSet result; 
  for(int i = 0; i < pC1.size; i++){
    for(int j = 0; j < pC2.size; j++){
      if(pC1.micro_context(i).dimension==pC2.micro_context(j).dimension){
        commonMC1.add(pC1.micro_context(i));
        commonMC2.add(pC2.micro_context(j));
        interMicroContext_i.add(pC1.micro_context(i));
        interMicroContext_i.add(pC2.micro_context(j));
        interMicroContext.add(interMicroContext_i);
        break;
      }
    }
  }
  //build commonCombination {[e:1, d:1],[e:1, d:2]...}
  //pointer for combining all the possible micro contexts in interMicroContext
  int iniposition=0; 
  //any simple context element of the context set commonCombination
  SimpleContext midReslt; 
  buildCommonCombination(interMicroContext, commonCombination, midResult, iniposition ); 
  uniqueMC1=getUniqueMCs(c1, commonMC1);
  uniqueMC2=getUniqueMCs(c2, commonMC2);
  //build the final result {[f:1, e:1, d:1, t:4]...}
  result=buildAllCombination(uniqueMC1,uniqueMC2,commonCombination);  
  return result;
}
\end{lstlisting}

\begin{lstlisting}[
    label={list:helpingMethodForUnionSimpleContext_buildCommonCombination},
    caption={Algorithm for implementing helping method \api{buildCommonCombination} for \api{union} on simple context},
    style=codeStyle
    ]
void buildCommonCombination(Vector pInterMicroContext, ContextSet result, SimpleContext pMidResult, int pPosition){ 
  //Passing by reference is used, thus void the return type
  if(pPosition==pInterMicroContext.size){
    //Finish one path of combination: eg.[e:1, d:2]
    result.add(pMidResult.clone()); 
    //Prepare to start another way of combination: 
    //eg. if pMidResult=[e:1, d:1], then after this, pMidResult=[e:1] 
    //waiting for the construction of pMidResult=[e:1, d:2]
    pMidResult.removeElement(pMidResult.lastElement()); 
    return;
  }
  else{ 
    //Constructing the possible combination
    Vector tempSC=pInterMicroContext.elementAt(position);
    position++;
    for(int i = 0; i < tempSC.size; i++){
      MicroContext tempMC = tempSC.elementAt(i);
      pMidResult.add(tempMC);
      //Call buildCommonCombination to finish one path of combination
      //eg: if pMidResult=[e:1], the call would add [d:1], 
      //thus making pMidResult=[e:1, d:1] 
      buildCommonCombination(pInterMicroContext, result, pMidResult, pPosition); 
    }
    if(pMidResult.size!=0){
      //If no micro context left, the recursive call ends.
      //Preparing for the next combination: 
      //eg. if pMidResult=[e:1], this operation clears it, 
      //waiting for the next path of [e:2,...]
      pMidResult.removeElement(pMidResult.lastElement());
      return;
    }
  }
}
\end{lstlisting}
\begin{lstlisting}[
    label={list:helpingMethodForSimpleUnionContext_buildAllCombination},
    caption={Algorithm for implementing helping method \api{buildAllCombination} for \api{union} on simple context},
    style=codeStyle
    ]
ContextSet buildAllCombination(Vector pUniqueMC1, Vector pUniqueMC2, ContextSet pCommonCombination){
  ContextSet result;
  for(int i = 0; i < pCommonCombination.size; i++){
    SimpleContext tempSC=pCommonCombination.simple_context(i);
    for(int p = 0; p < pUniqueMC1.size; p++){ 
      //eg. tempMC=[f:1]
      MicroContext tempMC=pUniqueMC1.elementAt(p);
      //insert [f:1] before [d:1, e:1] etc. 
      tempSC.insertElementAt(tempMC, p); 
    }
    for(int q = 0; q < pUniqueMC2.size; q++){
      //eg. tempMC=[t:4]
      MicroContext tempMC=pUniqueMC2.elementAt(q); 
      //append [t:4] after [d:1, e:1] etc.
      tempSC.add(tempMC); 
    }
    result.add(tempSC);
  }
  return result;
}
\end{lstlisting}

\begin{lstlisting}[
    label={list:helpingMethodForUnionSimpleContext_getRemainingUniqueMCs},
    caption={Algorithm for implementing helping method \api{getUniqueMCs} for \api{union} on simple context},
    style=codeStyle
    ]
Vector getUniqueMCs(SimpleContext pSC, Vector pMicroContext_p){
  Vector microContext_l;
  boolean picked=false;
  for(int p = 0; p < pSC.size; p++){
    MicroContext tempMC1=pSC.micro_context(p);
    for(int q = 0; q < pMicroContext_p.size; q++){
      MicroContext tempMC2=pMicroContext_p.elementAt(q);
      if(tempMC1 == tempMC2){
        picked=true;
        break;
      }
    }
    if(picked==false){
      microContext_l.add(tempMC1);
    }
    else
      picked=false;
   }
   return microContext_l;
}
\end{lstlisting}

\begin{lstlisting}[
    label={list:unionContextSet},
    caption={Algorithm for implementing \api{union} on context set},
    style=codeStyle
    ]
ContextSet union(ContextSet s1, ContextSet s2){
  DimensionSet interDimSet;
  for(int i = 0; i < s1.size; i++){
    for(int j = 0; j < s2.size; j++){
      for(int k = 0; k < s1.simple_context(i).size; k++){
        for(int l = 0; l < s2.simple_context(j).size; l++){
          if(s2.simple_context(j).micro_context(l)== s1.simple_context(i).micro_context(k))
            interDimSet.add(s1.simple_context(i).micro_context(k));
        }
      }
    }
  }
  ContextSet X1;
  for(int i = 0; i < s1.size; i++){
    for(int j = 0; j < s2.size; j++){
      X1.add(union(s1.simple_context(i), hiding(s2.simple_context(j), interDimSet)));
    }
  }
  ContextSet X2;
  for(int j = 0; j < s2.size; j++){
    for(int i = 0; i < s1.size; i++){
      X2.add(union(s2.simple_context(j), hiding(s1.simple_context(i), interDimSet)));
    }
  }
  for(int t = 0; t < X2.size; t++){
    X1.add(X2.simple_context(t));
  }
  X1.removeDuplicateContext();
  return X1;
}
\end{lstlisting}

\section{Implementation of Context Calculus in the GIPSY}

In order to execute a Lucid program, all the SIPL (Specific Intensional Programming Language)~\cite{aihuawu02,mokhovmcthesis05} ASTs
(abstract syntax tree) are translated into their {\gipl} counterparts using semantic translation rules establishing the specific-to-generic equivalence between the two languages~\cite{paquetThesis, aihuawu02, chunleiren02, mokhovmcthesis05}. The translated AST, together with the dictionary~\cite{aihuawu02, mokhovmcthesis05} are then fed to the Execution Engine, namely the GEE~\cite{paquetThesis,mokhovmcthesis05,bolu03} for the runtime execution. However, this translation approach cannot be easily adopted by {\lucx}. There's no such object as a {\it context} in {\gipl} and also the translation for context calculus operators would inevitably involve in recursive function calls, which are flattened before processing by the  {\gee}. As the notion of context is actually an essential concept for the {\lucid} programming language and we already have a type system in the GIPSY~\cite{gipsy-type-system-lpar08}, it is necessary and possible to keep the context as one of the GIPSY types in the type system. By defining this class, the context calculus operators can be implemented as member methods, which are going to be called at runtime by the {\gee} as it traverses the AST of {\lucx} and encounters those operators. In order to call those methods, the engine has to instantiate the context objects first. As stated earlier, a context is a collection of $<\!\!dimension:tag\!\!>$ pairs. During the instantiation, a semantic checking must be performed to verify if the tag is within the valid range of the dimension tag set. Thus, in order to implement the context calculus operators, we first have to introduce
the tag set classes into the {\gipsy}.

\subsection{Adding Tag Set Types into the GIPSY Type System}

As stated earlier in \xs{tagsets}, there are four kinds of tag sets. They are organized as shown in \xf{fig:tagsettypes}. The \api{TagSet} class is an abstract class and it's the parent of all of tag set classes. It has several data fields to keep the general attributes of tag sets, and it also has place-holder methods for certain common operators among all the tag sets such as equality method \api{equals()} and the set inclusion method \api{isInTagSet()}. There is also a group of interfaces for keeping the type information, for example, the class \api{OrderedFiniteTagSet} should implement the \api{IOrdered} and \api{IFinite} interfaces. Such mechanism also provides the facility of adding and defining proper operators into the proper tag set classes. Such as \api{getNext(poTag)}, which takes a tag object as parameter and returns the next tag value in the dimension, should be valid only for ordered sets. Then only the tag set classes implement the ordered interface should give the concrete implementation for this method.

\begin{figure*}[htb!]
	\begin{centering}
	\includegraphics[width=\textwidth]{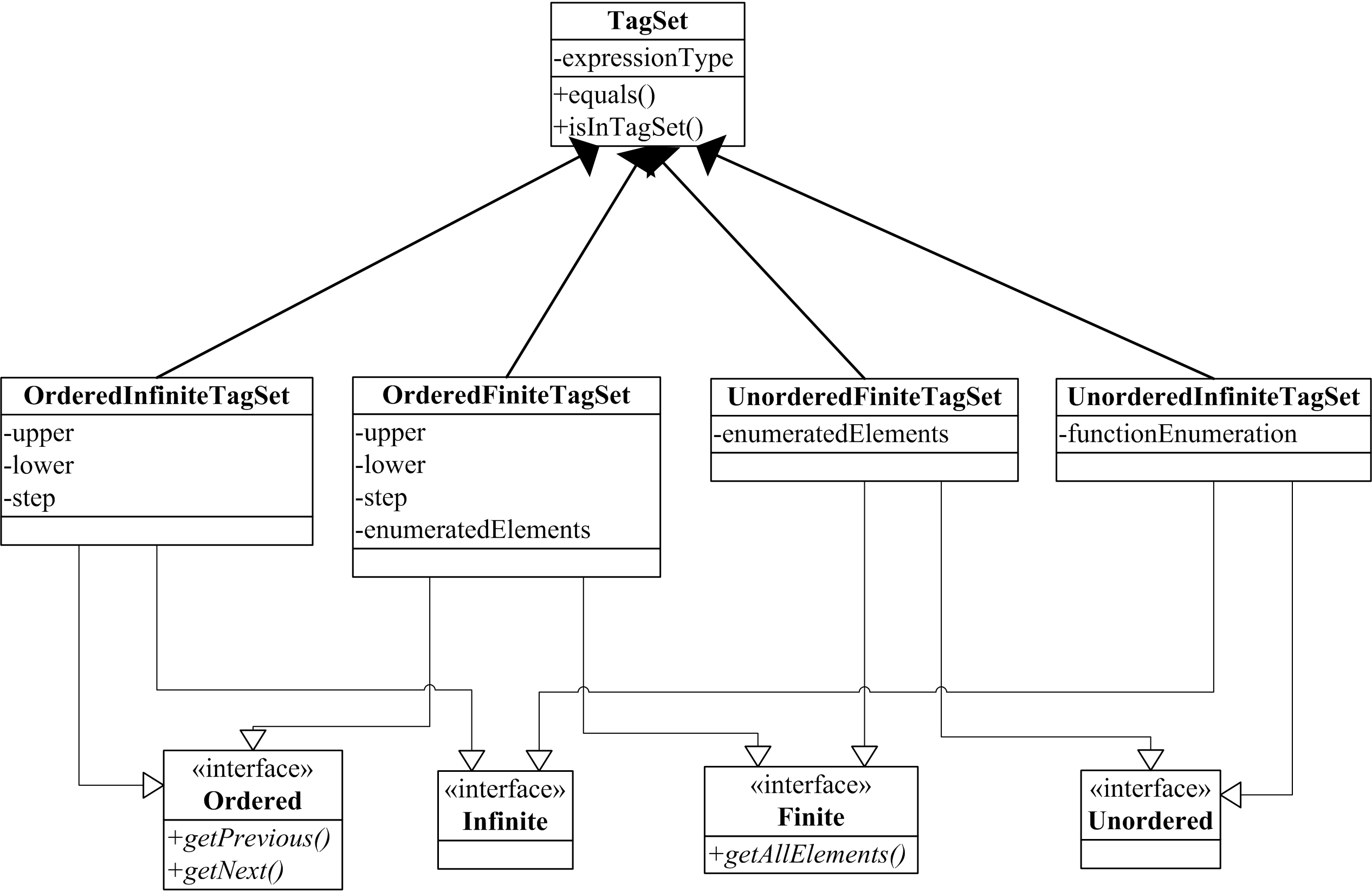}
	\caption{Tag Set Type Classes}
	\label{fig:tagsettypes}
	\end{centering}
\end{figure*}

\subsection{Adding Context into the GIPSY Type System}

As described in \xs{tagsets}, there are {\it simple context} and {\it context set} under the generic {\it context} type. We have \api{iContextType} data field in the \api{GIPSYContext} class to keep this information. \api{oSet} is the actual container of either micro contexts (for simple context) or simple contexts (for context set). The \api{Dimension} class has an object of type of \api{GIPSYIdentifier} called \api{oDimensionName} to specify its name and \api{oTagSet} to keep the information of the tag set attached to it. It also has a reference \api{oCurrentTag}, which is set to the current tag value inside the dimension, by adding this field, the notion of micro context can be expressed, since micro context is nothing but a pair of $<\!\!dimension:tag\!\!>$ and if we introduce another type of micro context, there would be data redundancy because it is going to be used only when constructing a simple context. Therefore, to sum up, a simple context is represented by a collection of \api{Dimension} with \api{oCurrentTag} specified and a context set is represented by a collection of \api{GIPSYContext} objects, with the \api{iContextType} set to \api{SIMPLECONTEXT}. \xf{fig:GIPSYContext} shows the structure of the \api{GIPSYContext} and related classes.

\begin{figure}[htb!]
	\begin{centering}
	\includegraphics[width=.5\textwidth]{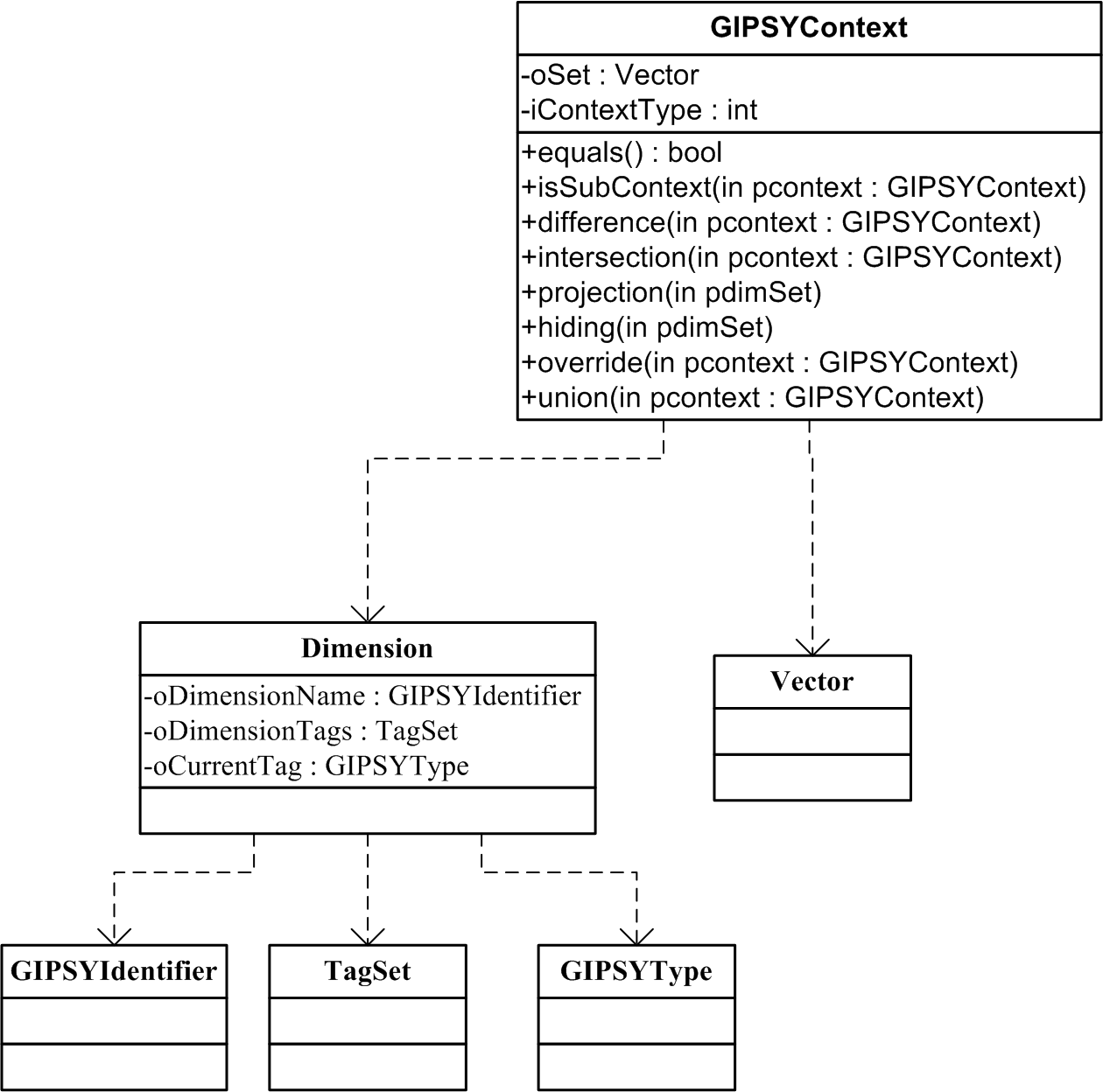}
	\caption{GIPSYContext class}
	\label{fig:GIPSYContext}
	\end{centering}
\end{figure}

\subsection{Semantic Checking for Context Type}

The GIPSY is equipped with both static (compile-time) and dynamic (run-time) type checking mechanisms. With the addition of the above-defined \api{GIPSYContext}, \api{Dimension} and \api{TagSet} classes, the existing static and dynamic semantic checkers are extended in the occurence of these types being computed by the compiled/executed {\lucx} programs. The sections that follow provide discussions related to the introduction of such static/dynamic semantic checks in the compiler and run-time system. 

\subsubsection{Validity of Tag Value Inside a Context}

A context is a relation between a dimension and a tag. When a context expression is specified, it always contains one or more \verb|[dimension:tag]| pairs. As the tag is the index of a dimension to mark a particular position for evaluation, it is necessary to check first if the \verb|tag| is part of the valid tags for this \verb|dimension|, in other words, that it is an element of the tag set attached to this dimension. This can be resolved by calling the set inclusion method defined for each tag set. When the tag expression is simply a constant or a literal, this checking is performed at compile time by traversing the AST and calling the set inclusion method. When the tag expression is complex, the semantic checking should be delayed to runtime by the execution engine to compute the resulting values and subsequently do the semantic checking when it tries to instantiate the corresponding \api{GIPSYContext} object.

\subsubsection{Validity of Operands for Context Calculus Operators}

As defined earlier, the context calculus operators have some semantic restrictions on what are the valid operands, such as the \api{union} operator requires its operands either to be both of {\it simple contexts} or both of {\it context sets}. When the tag expressions are constants or literals, such checking is to be performed at compile time by traversing the AST and get the type of contexts. If the tag expressions are complex, this checking is deferred to runtime by the engine.

\section{Conclusion}

By introducing contexts as first-class values, a set of context calculus operators are allowed to be performed on the context objects to provide us the facility of constructing and manipulating contexts in different application domains in the {\gipsy}. As we abstract the context into an object, the essential relation of dimension and tag is also properly and more completely defined by introducing tag set types. Since we have the GIPSY type system containing all the possible data types in {\lucid}, context, as one of the first class objects, is taken as a standard member of the type system. By giving the Java class representation for context, the context calculus operators have been implemented as member functions inside the \api{GIPSYContext} class.

\section{Future Work} 

The context calculus operators implemented in the \api{GIPSYContext} class have already been fully tested using JUnit ~\cite{junit}. The next step is to make them completely executable at run-time on the {\gee} side. The {\gee} evaluates Lucid expressions by traversing the ASTs provided by the compilers. Thus, in order to compute the context calculus, we have to make the $context$ and $context\; calculus$ nodes are recognizable by the engine. When a $context\; calculus$ node is encountered, it can be evaluated by instantiating a context object and calling the member function defined.

%

%
\label{sect:bib}
\bibliographystyle{abbrv}
\bibliography{gipsy-context-calculus-09}

\end{document}